\documentclass[12pt]{article}

\usepackage{epsfig}

\begin{document}

\bibliographystyle{unsrt}

\begin{flushright}
IFT 2001/39
\end{flushright}
\begin{center}
\vspace{0.6cm}
\Large{Heavy Quark Production at the TESLA Collider and its Sensitivity
to the Gluon Content in Photon}\\
\vspace{1cm}
\large{P. Jankowski and M. Krawczyk}\\
\vspace{0.4cm}
{\sl Institute of Theoretical Physics, Warsaw University}\\
{\sl ul. Ho\.za 69, 00-681 Warsaw, Poland}\\
\vspace{0.6cm}
\large{A. De Roeck}\\
\vspace{0.4cm}
{\sl CERN, 1211 Geneva 23, Switzerland}
\vspace{0.5cm}
\end{center}

\begin{abstract}
Heavy quark production is studied at the high energy linear $e^+e^-$ collider 
(LC) TESLA both in its nominal and Photon Collider (PC) mode. Leading order 
cross-sections are calculated for the production of heavy quarks, 
$e^+e^- \rightarrow e^+e^-\: Q(\bar{Q}) X$, at high transverse momenta.
The sensitivity of this process to the gluon content in the photon is studied.
\end{abstract}

\vspace{1.2cm}

An $e^+e^-$ linear collider with a centre of mass system (CMS) energy in the 
range of 500 to 1000 GeV is the prime candidate for the next high energy 
accelerator project after the LHC. Recently TESLA~\cite{TDR} has completed a 
Technical Design Report for such a collider. For the study of a future 
electron-positron Linear Collider (LC) it is important to examine the physics 
potential of its main and possible additional options. Such an option is the 
Photon (or Compton) Collider (PC) in which high energy real photons can be 
obtained by backscattering photons from a laser beam on the electron or 
positron beam \cite{GINZ,TEL}. This way an excellent tool for the study of 
$\gamma \gamma$ collisions at high energies can be provided.

\bigskip

In high energy $e^+e^-$ collisions the hadronic final state is predominantly 
produced in $\gamma^{\star}\gamma^{\star}$ interactions where the virtual 
photons ($\gamma^{\star}$) are almost on mass shell. Their scattering can be 
effectively described with the interaction of real photons with energy spectra
given by the Weizs\"{a}cker-Williams (WW) approximation~\cite{ww}. Such a 
spectrum is generally quite soft, as shown in Fig.~1. In contrast, a Photon 
Collider, based on Compton scattering, provides beams of real photons which 
are much harder than those from a WW distribution (see Fig.~1). Furthermore 
these photons can be produced in a definite polarization state. While the 
energy spectrum of the PC at the high energy end can be well calculated based 
on a few parameters of the laser and incident electron beam, the lower energy 
end depends much more on the details of machine and operation. Therefore we 
limit ourself to study  only  the high energy peak of 
$0.6<y = E_{\gamma}/E_e <0.83$ ~\cite{TDRP}. The spectra used in this paper 
are described in the Appendix.

\bigskip

The main goal of this study is to compare the LC and PC potential for probing 
the gluon distribution in the photon via heavy quark production. This analysis
does not make use of polarization. The measurement of the process 
$e^+e^-\rightarrow e^+e^-Q (\bar Q) X$ is an important test of QCD. It is well 
known that this process is promising for probing the structure of the photon,
see e.g.~\cite{DREES, CGKKKS} and \cite{DGP}, where a related topic is 
considered. Heavy quarks can be produced in $\gamma \gamma$ collisions through
three mechanisms. Direct (DD) production occurs when both photons couple 
directly to a $Q\bar{Q}$ pair. In single resolved photoproduction processes 
(DR) one of photons interacts via its partons with the second photon. When 
both photons split into a flux of quarks and gluons, the process is labelled a
double resolved photon (RR) process.

\bigskip

There are various schemes proposed for calculating cross-sections for processes
involving heavy quarks, for the latest overview see ref \cite{TUNG2001}. 
Two standard schemes will be used in this analysis \footnote{a recent attempt
to construct a new composite scheme can be found in \cite{TUNG2001}}.
They differ by the number of quark flavours which correspond to partons of the
photon (active quarks). In the case of the massive quark scenario, the so 
called Fixed Flavour Number Scheme (FFNS), the photon ``consists'' only of 
light quarks and gluons, which may interact, and massive heavy quarks can be 
only created in a hard process e.g. via gluon-gluon fusion. The massless quark
scheme, Zero-mass Variable Flavour Number Scheme (ZVFNS), considers apart from
the gluons and $u, d$ and $s$ quarks also heavy quarks as active quarks, which
are all treated massless. All the partonic reactions in both schemes 
contributing in the LO calculation to heavy quark production are shown in 
table 1. Present available data on charm quark production in $\gamma \gamma$ 
processes is still not precise enough to allow to discriminate between the
predictions of these two production schemes. Besides, these theoretical 
approaches are expected to be in turn appropriate for describing different 
ranges of the hard scale ($\mu$) of the process: The massive and massless 
schemes should be applied below ($\mu\leq m_Q$) and above ($\mu\gg m_Q$)
the mass scale, respectively. For the charm quark and for transverse momentum 
$p_T$  of the partons larger than e.g. 10 GeV, we expect to be closer to the 
region of applicability of the ZVFNS scheme than the FFNS scheme. For 
completeness we performed the calculation in both schemes which can be treated
as an estimate of the theoretical uncertainty.

\bigskip

We calculate the LO QCD cross-sections for $c$ and $b$ quarks produced with 
large $p_T$ at the $e^+e^-$ LC colliders at energies of 300, 500 and 800 GeV,
respectively, and at the PC based on the corresponding ($e^+e^-$) LC with 
these energies. For comparison we study also the corresponding cross-sections
for the LEP collider at 180 GeV. To test the sensitivity of the considered 
processes to the gluonic content of the photon we use two different parton 
parametrizations for the real photon: GRV \cite{GRV} and SaS1d \cite{SaS}. 
We choose the GRV('92) parametrization because in its construction heavy quarks
are treated as massless above the heavy quark threshold region ($W \gg 2m_Q$),
and moreover recent experimental results \cite{Adloff} indicate that GRV gluon
distribution is close to the data. The choice of the SaS1d parametrization is 
motivated by the fact that it provides results in agreement with expectations
of Gribov factorization and the QCD sum rule for the structure of the 
photon~\cite{GrSum}. Both these parametrizations are based on QCD fits to 
photon structure function data measured in $e\gamma$ collisions at $e^+e^-$ 
experiments \cite{NISIUS,KSZ}. They have different assumptions for the gluon 
content, which is only weakly constrained by the data. The GRV and SAS1d 
distributions both start the evolution from a small scale, $\mu^2_0 = 0.25$ GeV
$^2$ and 0.36 GeV$^2$, respectively, and consequently both parton densities 
predict a rise of the gluon density at small $x$. Different treatment of 
valence quark distributions in the vector mesons leads to a larger gluon 
component at small-$x_{\gamma}$ of the photon for the GRV set compared to the 
SAS1d one.

\bigskip

As was mentioned above, in the massive (FFNS) calculations the number of active
flavours ($N_f$) is taken to be 3. In the massless (ZVFNS) scheme it varies 
from 3 to 5 depending on the value of the  hard (factorization, 
renormalization) scale $\mu$. The  $c$ ($b$) quarks are included in the 
computation provided that $\mu > m_c$ ($m_b$) with $m_c=1.5$ GeV ($m_b=4.5$ 
GeV). When charm quark production is calculated in the ZVFNS scheme the bottom
quarks are always excluded, hence $N_f = 3$ or $N_f = 4$. Also the QCD energy 
scale $\Lambda_{QCD}$, which appears in the one loop formula for the strong 
coupling constant $\alpha_{s}$, depends on the number of active flavours. We 
take this scale, denoted as  $\Lambda_{QCD}^{N_f}$, to be:
\begin{equation}
\Lambda_{QCD}^{3} = 232 ~~~~\Lambda_{QCD}^{4} = 200 ~~~~\Lambda_{QCD}^{5} 
= 153 \:\rm MeV
\end{equation}
as in \cite{GRV}.
The hard scale in the calculation of the cross-section $\mu$ is taken to be 
the transverse mass of the produced heavy quark $m_T=\sqrt{m_Q^2+p_T^2}$.

\bigskip

In the calculation we include  direct (DD), single (DR) and double resolved
(RR) photon processes. We study {\large $\frac{d^2\sigma}{dp_T^2d \eta}$}, 
with $\eta = ${$\huge \frac{1}{2}\ln \frac{E-p_L}{E+p_L} $}  the rapidity 
of the produced heavy quark and $p_T$ its transverse momentum. The results for
the charm quark production presented in Figs.~2 and ~3 were obtained in the 
ZVFNS scheme  for both types of initial photon spectra. The calculation was 
performed using the GRV parton parametrization. In Fig.~2 a fixed energy 
$\sqrt s$=500 GeV and a $p_T$ =10 GeV for the charm quark was assumed.
Both resolved photon contributions, DR and RR, to the process $e^+e^-
\rightarrow e^+e^-Q (\bar Q) X$  are dominated by reactions initiated by 
gluons, especially for a PC  at high $\eta$ values. The dominance of the gluon
initiated over the quark initiated processes is larger for the PC spectrum; it
is  also larger in the FFNS scheme (not shown) compared to the ZVFNS one.

\bigskip

Fig.~3 shows a comparison of the direct (DD), single resolved (DR), 
double resolved (RR) contributions and their sum for WW and LASER spectra
for various CMS energies of the $e^+e^-$ colliders, for charm quarks with a 
$p_T$ of 10 GeV. An interesting pattern is observed. In the case of
an $e^+e^-$ LC with a WW  photon spectrum  either the direct production
of $c$  dominates, or resolved and direct photon contributions are found to be
of the same size. The DR and RR contributions increase with increasing CMS 
energy of LC. Nevertheless in the considered range of the LC energies, they 
do not play a dominant role for heavy quark \footnote{A similar effect is 
observed for $b$ quark  production (not shown)} production
in the $e^+e^-$ mode. The contrary is observed  for a PC: charm quark 
production is always dominated by resolved photon interactions. The direct 
contribution is limited to  high absolute values of the rapidity distribution.
For a $p_T=10$ GeV this rapidity region is outside the range of $|\eta|<2$ 
which could be covered by a typical detector. Lowering the $p_T$ of the 
observed charm quarks amplifies the contribution of the resolved process to 
the total cross-section. The direct contribution becomes even less important 
with increasing CMS  energy. Thus the charm production cross-section is much 
more sensitive to the parton  distribution of the photon for a PC compared to 
a LC. Since for the PC option the resolved photon contributions are dominated 
by the processes involving gluons (Fig.~2) this offers an excellent tool for 
measuring the gluonic content of the photon. We note here also that an 
extention of the  experimental reach to observe charm quarks at rapidities 
beyond $|\eta|=2$ would allow to access a kinematic range where the gluon 
contribution dominates most strongly over other resolved photon contributions.

\bigskip

For the ZVFNS scheme and the PC option we find that the processes initiated by
$c$ quarks constitute about 52-67\% of the total cross-sections presented in 
Fig.~3, and about 99\% of the contribution which arises due to the processes 
initiated by quarks (not shown). Since the heavy quark content of the photon 
is closely related to the gluon density in the photon, the heavy quark 
production discussed in the paper exhibits also sensitivity to the gluonic 
density via the heavy quark content of the photon \cite{kraemer}. Since for 
the LC option the cross-section has a large contribution from the direct 
process, this effect should be most visible in data from a PC, which are 
dominated by the resolved photon processes instead.

\bigskip

An important feature of the results for cross-section seen in Fig.~3 is the 
increase of the resolved photon process contributions, and consequently also an
growing sensitivity to gluons, with increasing CMS energy. This arises from the
fact that at higher energies we probe regions of small Bjorken-$x_{\gamma}$ 
values. The minimal $x_{\gamma}$ value that could be reached in LC for 
$p_T=10$ GeV varies from $\sim 0.004$ for $\sqrt{s}=300$ GeV to $\sim 0.0006$ 
for $\sqrt{s}=800$ GeV ($\sim 0.01$ in LEP). The detector cut $|\eta|<2$ 
applied in this analysis increases those values to $\sim 0.006$ and 
$\sim 0.0018$, respectively. An extra cut on the LASER photon spectrum 
($0.6<y<0.83$) in the PC mode shrinks the $x_{\gamma}$ accesible range even 
further. The numbers read then: $\sim 0.008$ for $\sqrt{s}=300$ GeV and 
$\sim 0.0023$ for $\sqrt{s}=800$ GeV. Therefore the RR contribution grows 
weaker with the CMS energy in the PC case comparing to the LC case. For the DR
case in the PC mode the $0.6<y<0.83$ cut puts an additional, kinematic 
constraint. Since one of the photons interacts directly with 
$E_{\gamma}>0.6E_e$ the energy of the produced heavy quarks must grow with 
$E_e$. If we choose them to have relatively small $p_T$ equal to 10 GeV it is 
more difficult to produce them in the central $\eta$ region where 
$E\approx p_T$. Hence instead of the increase with the energy we see a slight 
decrease of the DR contribution around $\eta=0$. Due to the same kinematic 
effect DD events cannot be produced close to $\eta = 0$. We note as above that
reaching in experiment values of rapidity higher than $|\eta|>2$ would be of 
high importance. We then could explore kinematic region of strongest resolved 
contributions to the cross-sections. 

\bigskip

The results discussed so far are not significantly affected by the choice of 
scheme for the heavy quark calculation as is shown in Fig.~4 for the $c$-quark
rapidity distribution: for both the FFNS and ZVFNS scheme the cross-section is
much larger for the PC than for the LC case. 

\bigskip

More details of the comparison of the FFNS and ZVFNS scheme results can be 
found in Figs.~5-9. The importance of the low $x_{\gamma}$ contributions is 
demonstrated  in Fig.~5, presenting the {\large $\frac{d\sigma}{dx_{\gamma}}$}
distributions of the charm production at a PC with $\sqrt{s_{ee}}=500$ GeV
(for the GRV parton parametrization), on a linear and logarithmic 
$x_{\gamma}$ scale. The cross-section has been integrated over the region 
$p_T>10$ GeV and $|\eta|<2$, realistic for a possible detector at a PC. 
The direct contribution is included into the bin with the highest 
$x_{\gamma}$. The double resolved part is counted for twice, once for each 
$x_{\gamma}$ value. The DR contributions are peaked  at small $x_{\gamma}$, 
while the RR ones have a flatter distribution. The expected differences 
between both schemes used in the calculation are visible: the ZVFNS scheme 
gives a larger contribution of DR and RR processes, especially for 
$x_{\gamma}>$0.2. The maximum in the DR distribution occurs at $x_{\gamma}$
close to 0.01, independently on the scheme. The RR process have a broad 
distribution between $x_{\gamma} \sim 0.03-0.05$ and 1.0, and the ZVFNS scheme 
result is larger  by more than one order of magnitude compared to the FFNS one
in this whole region. 

\bigskip

The sensitivity of the charm quark production process to the gluon distribution
is studied further by comparing the predictions obtained using two different 
parton parametrizations for the photon. In Fig.~6 (7) the ratio of the 
relative difference of the cross-sections 
{\large $\frac{d^2\sigma}{dp_T^2d \eta}$} is presented, obtained using the GRV
and SaS1d parton parametrizations in the ZVFNS (FFNS) scheme. As expected the 
cross-section obtained using the PC photon spectrum (LASER) leads to a larger 
sensitivity than the WW spectrum for a given energy of the $e^+e^-$ collider: 
the difference between results of these two parametrizations shown is 5-20\% 
for a WW and 30-50\% for LASER photon spectrum. Hence a PC collider has a 
larger sensitivity for testing the photon structure and especially its gluon 
density. It is worth  mention  that these results indicate however that 
an initial study of the partonic structure of real photon can be already 
made with an $e^+e^-$  LC itself.

\bigskip

Next we show some detailed characteristics of charm quark production at a PC.
Fig.~8 and 9   show the relative difference of $x_{\gamma}$ distribution
calculated with the GRV and SaS1d parametrizations, for the {\large 
$\frac{d\sigma}{dx_{\gamma}}$} cross-sections introduced before, for 
$|\eta|<2$. Results for $p_T>10$ GeV are shown in Fig.~8 (for both linear (A) 
and logarithmic (B) scale) and for $p_T>5$ GeV in Fig.~9 (linear scale only).  

\bigskip

Table~2 shows in detail how the gluon initiated resolved processes dominate
the charm quark production. For the case of $\sqrt{s}=800$ GeV they 
contribute 76\% and 98\% of the integrated cross-section calculated in the 
FFNS and ZVFNS scheme respectively. Further details on contributions of the 
gluon and charm quark initiated processes can be found in Table~3.

\bigskip

Up to now we presented only the results for $c$ quark production. The 
corresponding $b$ quark production has all the features listed above though 
the difference between a PC and a LC was found to be smaller. All calculated 
cross-sections for bottom production are found to be even more sensitive to 
the shape of gluonic content of the photon than the corresponding ones  
for charm quark production, but the cross-sections are smaller.

Next we calculate the number of events with heavy quarks, $c$ and $b$, 
assuming an $e^+e^-$ integrated luminosity of 300 fb$^{-1}$, which could be 
achieved at a high luminosity LC with one year of running. The corresponding 
luminosity of the PC for the region $0.6 < y < 0.83$ is about a factor 3 to 4 
lower~\cite{ggpaper}. In the FFNS scheme the event rates for charm quark 
production at a LC for the different CMS energies 300-800 GeV amount to 
$1-4\cdot 10^{6}$ events for $p_T>$ 10 GeV; the corresponding numbers at a PC 
are $9-25\cdot 10^6$ events. Hence these calculations predict a high number of
the heavy quarks, $c$ and $b$, produced at the considered CMS energies of the 
$e^+e^-$ collider of 300, 500 and 800 GeV for  both the LC and PC options.

In practice, charm quarks with a $p_T> 5 $ GeV produced e.g. via $D^*$ decays 
can be detected in a generic LC detector without dedicated detectors in the 
rapidity range of $|\eta| < 1.5-2$. This restricts the lowest reachable 
$x_{\gamma}$ values to 0.0011 and 0.0023 for a minimum $p_T$ of 5 and 10 GeV, 
respectively, for a PC collider at 800 GeV. The cross-sections for $c$ and $b$
production in this phase space region are presented in Table 4. The 
corresponding event rates (for FFNS scheme leading to smaller rates) are given
in Table 5. It shows that the event rate at a PC is still larger than the one 
at a LC, but the difference is reduced to a factor 2 to 4. Hence some of the 
advantage of the PC over the LC is lost due to the charm quark production at 
large $\eta$ in case of the PC (see Fig.~4), which will go undetected with the
presently planned detectors.

To measure the charm content of the photon e.g. $D^*$ mesons can be selected 
to reconstruct the kinematics of the scattering process. The efficiency   
including fragmentation fraction and branching ratios, is typically around a 
few times 10$^{-3}$~\cite{OPALF2C}. Hence the number of reconstructed 
events per year with charm quark with  $p_T> 10 $ GeV will be approximately a 
few thousand for the LC and about ten thousand events for the PC. The numbers 
are about 5 times larger for a reduced cut of $p_T>5$ GeV. These data will 
clearly allow for precision measurements of charm quark production and further
on the gluon distribution in the photon. The statistical precision of the 
measurements of {\large $\frac{d^2\sigma}{dp_T^2d \eta}$} with e.g. ten bins  
will be approximately 5\% at the LC and a few \% at the PC. The event rates in
the ZVFNS scheme can be larger by a factor 3, compared to the FFNS scheme.

\bigskip 
\bigskip

In conclusion, the calculated cross-sections for heavy quarks ($c$ and $b$) 
production in two photon collisions show a  higher sensitivity to the 
parton distributions of the photon in case 
of a Photon Collider compared to a $e^+e^-$  Linear Collider. This does not 
depend on the particular scheme used to calculate the heavy quark 
cross-sections. Since the resolved photon contribution is to a large extent 
dominated by gluon induced processes, especially for a high energy PC, we 
conclude that heavy quark production provides indeed a sensitive probe of 
the gluon content of the photon. Combining the above features with the 
larger cross-sections achieved at energies of the $e^+e^-$ collisions at a PC 
favours this option for future photon structure research. A high luminosity 
$e^+e^-$ collider which drives the PC collider will however be essential.

\bigskip

\begin{flushleft}
{\bf \large Appendix}
\end{flushleft}

The following formula of the Equivalent Photon Approximation is used:
\begin{eqnarray}
f_{\gamma}(y) = \frac{\alpha}{2\pi}( \frac{(1+(1-y)^2)}{y}
                                      \log(\frac{Q^2_{max}}{Q^2_{min}} ) -
            \frac{2(1-y)}{y}(1-\frac{Q^2_{min}}{Q^2_{max}}) ) \nonumber \\
Q^2_{min} = \frac{m_e^2 y^2}{1-y} \\
Q^2_{max} = 1 {\textrm GeV} ^2 \nonumber
\end{eqnarray}
where $y = \frac{E_{\gamma}}{E_e}$ and $m_e$ is the mass of the electron.

In case of the PC mode we use the original energy spectrum of 
unpolarized photons \cite{GINZ}:
\begin{eqnarray}
f_{\gamma}(y) = \frac{1}{\sigma_c^{np}}[\frac{1}{1-y}+1-y-4r(1-r)], 
 \nonumber \\
\sigma_c^{np} = (1-\frac{4}{\kappa}-\frac{8}{\kappa^2})\ln(\kappa+1) + 
\frac{1}{2}+\frac{8}{\kappa}
-\frac{1}{2(\kappa+1)^2}, \\
r = \frac{y}{\kappa(1-y)}, \nonumber
\end{eqnarray}
where $\kappa$ is a parameter giving the restriction of $y$ value:  
$y${\large $<\frac{\kappa}{1+\kappa}$}. It is argued \cite{TEL} that the 
optimal value of $\kappa$ is 4.83, which gives a cut of $y$, 
$y_{max}=0.83$. We have chosen these values for this analysis.
Because the part of the spectrum with $y < 0.6$ is very 
sensitive to the technical parameters of the PC such as the size of the beam
we used only its high energy peak of $0.6<y<0.83$ \cite{TDRP}.

\bigskip

\begin{flushleft}
{\bf \large Acknowledgment}
\end{flushleft}

M.K. has been partly supported by the Polish State Committee for 
Scientific Research 5 P03B12120 (2001) and European Commission 
5th framework contract HPRN-CT-2000-00149 

\clearpage

\vspace{2cm}

\begin{table}[htb]
\begin{center}
\begin{tabular}{|c|c|c|c|}
\hline
     &    DD    &    DR    &    RR    \\
\hline
FFNS & $\gamma+\gamma \to Q+\bar{Q}$ & $g+\gamma \to Q+\bar{Q}$ & $g+g \to Q+\bar{Q}$ \\
     &                               &                          & $q+\bar{q} \to Q+\bar{Q}$ \\
\hline
ZVFNS & $\gamma+\gamma \to Q+\bar{Q}$ & $g+\gamma \to Q+\bar{Q}$ & $g+g \to Q+\bar{Q}$ \\
     &                               & $Q(\bar{Q})+\gamma \to Q(\bar{Q})+g$ 
     & $q+\bar{q} \to Q+\bar{Q}$ \\
 &&  & $Q+\bar{Q} \to Q+\bar{Q}$ \\
 &&  & $Q+Q(\bar{Q}+\bar{Q}) \to Q+Q(\bar{Q}+\bar{Q})$ \\
 &&  & $Q(\bar{Q})+q(\bar{q}) \to Q(\bar{Q})+q(\bar{q})$ \\
 &&  & $Q(\bar{Q})+g \to Q(\bar{Q})+g$ \\
\hline
\end{tabular}
\end{center}
\caption{Parton reactions contributing to heavy quark production in 
$\gamma \gamma$ collision in the massive (FFNS) and massless (ZVFNS) schemes.}
\end{table}

\begin{figure}[h]
\vskip -5cm
\hskip -2cm
\psfig{file=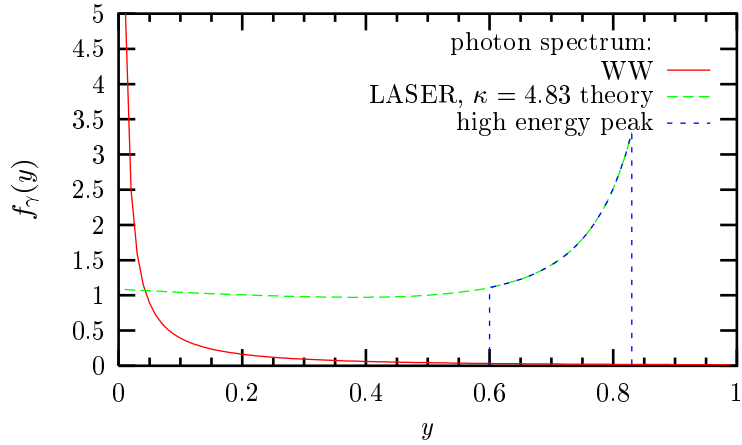}
\vskip -19cm
\caption{The photon spectra $f_{\gamma} (y)$ as a function of the energy 
fraction of the photon. A comparison is shown of the spectrum of photons  
radiated by the initial electron according to the Weizs\"acker-Williams (WW) 
approximation with the spectrum of (unpolarized) photons produced in the 
backscattering Compton process on the laser beam (LASER) for $\kappa=4.83$, 
see Appendix for definitions. For the LASER case the high energy peak 
$y > $  0.6 is indicated.}
\end{figure}

\clearpage

\begin{figure}[hb]
\vskip -4cm
\hskip -4cm
\psfig{file=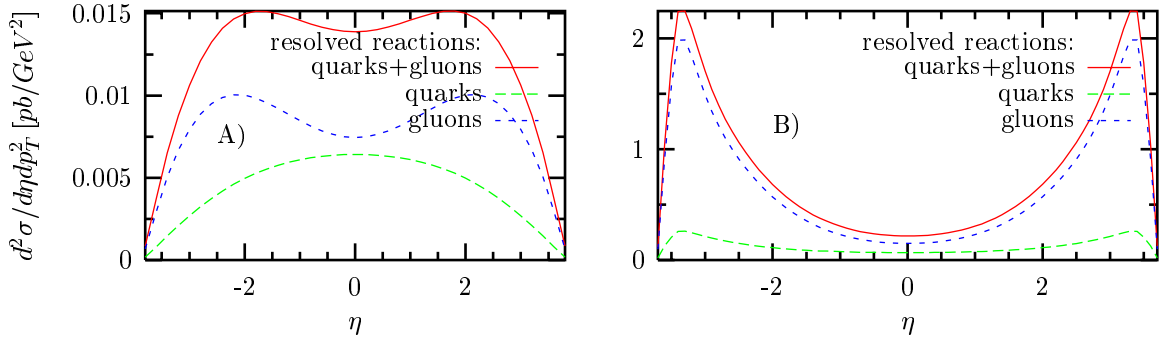}
\vskip -20.5cm
\caption{Comparison of parton processes  which involve  gluons with other 
partonic  contributions to the resolved (DR+RR) part of the cross-section 
{\Large $\frac{d^2\sigma}{dp^2_Td\eta}$} 
$e^+e^- \rightarrow e^+e^- c/\bar{c} X$ for $\sqrt{s}=500$ GeV. Results are 
obtained for charm quarks with transverse momentum $p_T=10$ GeV as a function 
of rapidity $\eta$ (in the $e^+e^-$ CM frame) in the ZVFNS scheme with the GRV
parton parametrization using: A) the WW  spectrum, B) the LASER  spectrum.}
\end{figure}

\clearpage

\begin{figure}[]
\vskip -5cm
\hskip -4cm
\psfig{file=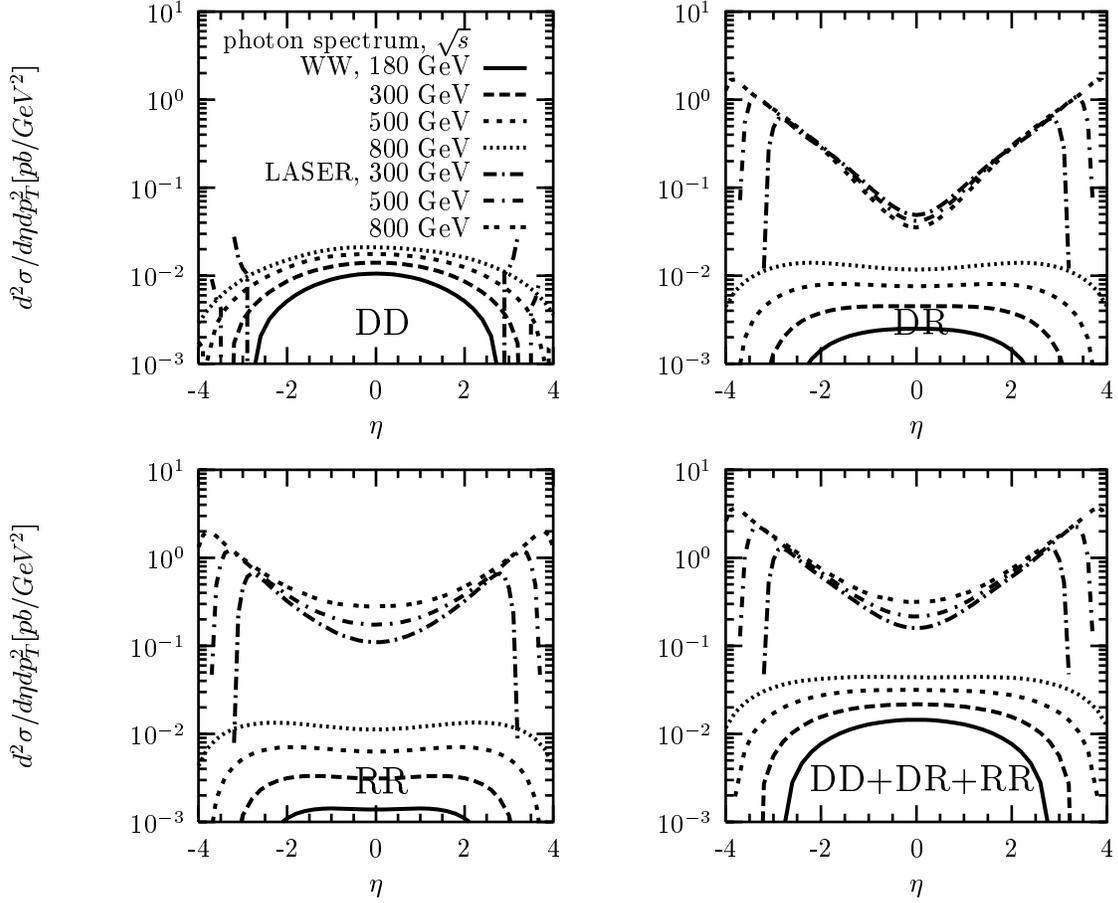}
\vskip -13.7cm
\caption{The cross-sections {\Large $\frac{d^2\sigma}{dp^2_Td\eta}$}
$(e^+e^- \rightarrow e^+e^- c/\bar{c} X)$ at various LC energies 
as a function of the $c/\bar{c}$ rapidity $\eta$ in the $e^+e^-$ CM frame 
at $p_T=10$ GeV presented  for two types of photon spectra (WW and LASER).
Separately the DD, DR, RR and their sum are shown. For a comparison also  
estimates for LEP 2 (WW, $\sqrt{s}$=180 GeV) are given. 
The GRV parton parametrization and the ZVFNS scheme are used.}
\end{figure}

\begin{figure}
\vskip -4.5cm
\hskip -2cm
\psfig{file=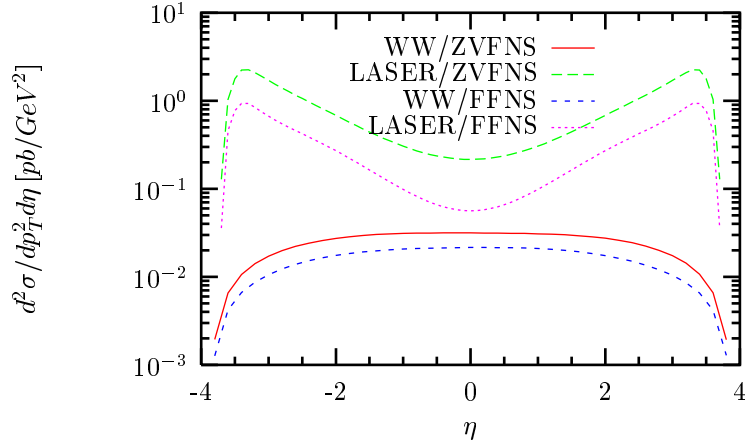}
\vskip -19cm
\caption{
The cross-section {\Large $\frac{d^2\sigma}{dp^2_Td\eta}$}
$(e^+e^- \rightarrow e^+e^- c/\bar{c} X)$ at $\sqrt{s}=500$ GeV as a function 
of $\eta$ for $p_T=10$ GeV. In the calculation the GRV parton parametrization 
is used. A comparison is shown between ZVFNS and FFNS schemes, and between the
WW and LASER photon spectra.}
\end{figure}

\begin{figure}
\hskip -2cm
\psfig{file=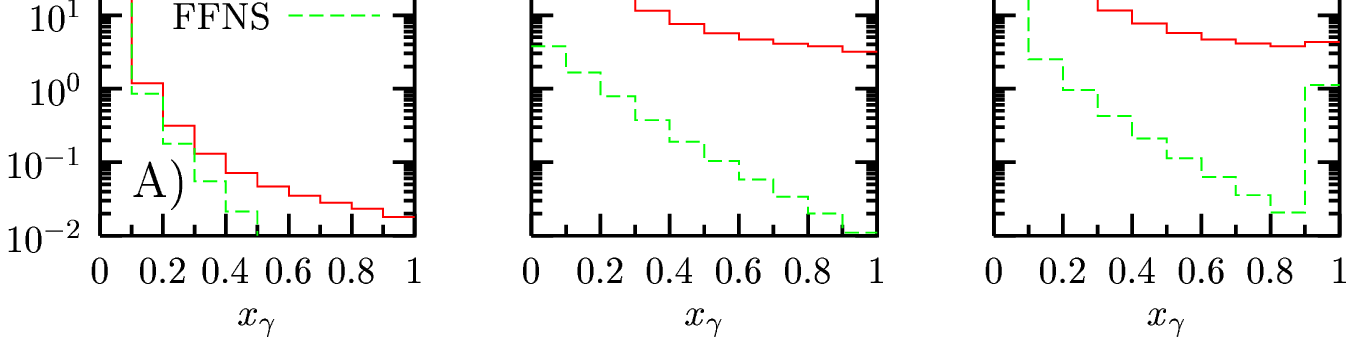}
\vskip -25cm
\hskip -2cm
\psfig{file=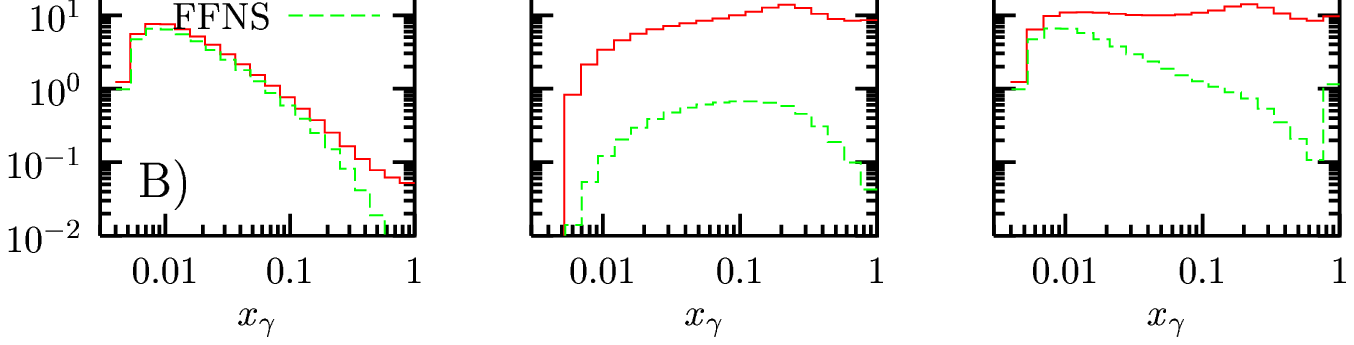}
\vskip -24cm
\caption{
The cross-section {\Large $\frac{d\sigma}{dx_{\gamma}}$} 
$(e^+e^- \rightarrow e^+e^- c/\bar{c} X)$ as a function of $x_{\gamma}$ for 
$\sqrt{s}=500$ GeV, integrated over $p_T>10$ GeV and $|\eta|<2$ for two types 
of schemes, ZVFNS and FFNS. The LASER spectrum and the GRV parton 
parametrization are used; A) linear, B) logarithmic $x_{\gamma}$ scale.}
\end{figure}

\clearpage

\begin{figure}
\vskip -5cm
\hskip -4cm
\psfig{file=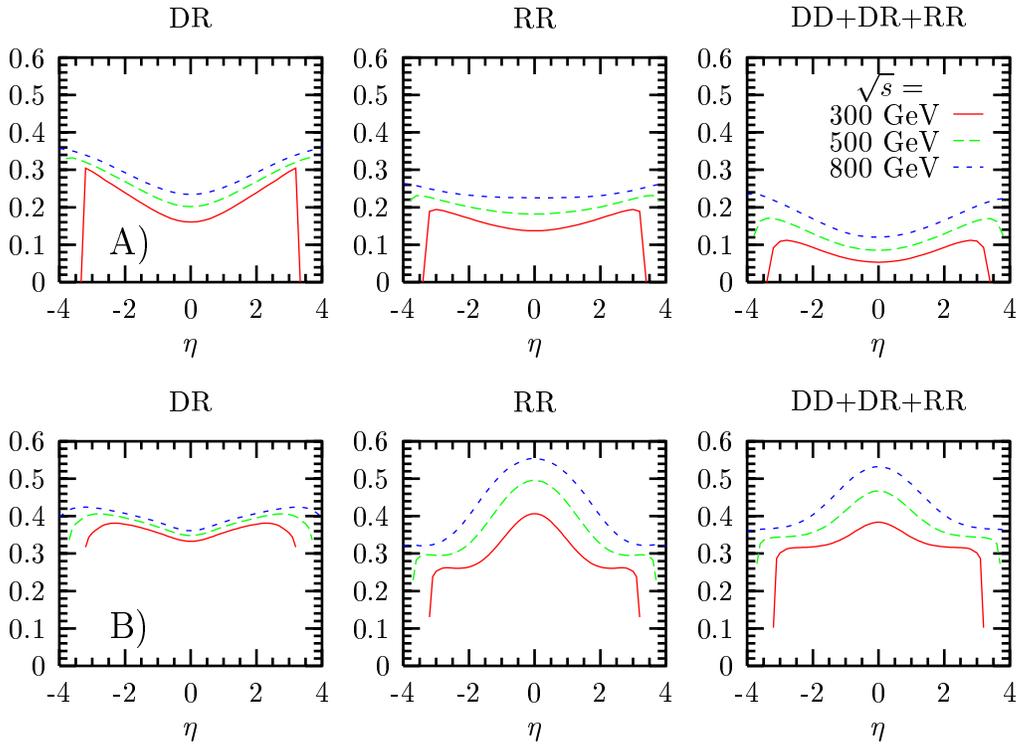}
\vskip -17cm
\caption{The ratio {\large $\frac{GRV-SAS1d}{GRV}$} of the cross-section 
{\Large $\frac{d^2\sigma}{dp^2_Td\eta}$}
$(e^+e^- \rightarrow e^+e^- c/\bar{c} X)$ as a function of $\eta$ for 
various LC energies and for $p_T=10$ GeV. Results  are obtained in the 
ZVFNS scheme for two photon spectra: A) WW, B) LASER.}
\end{figure}

\begin{figure}
\vskip -4cm
\hskip -4cm
\psfig{file=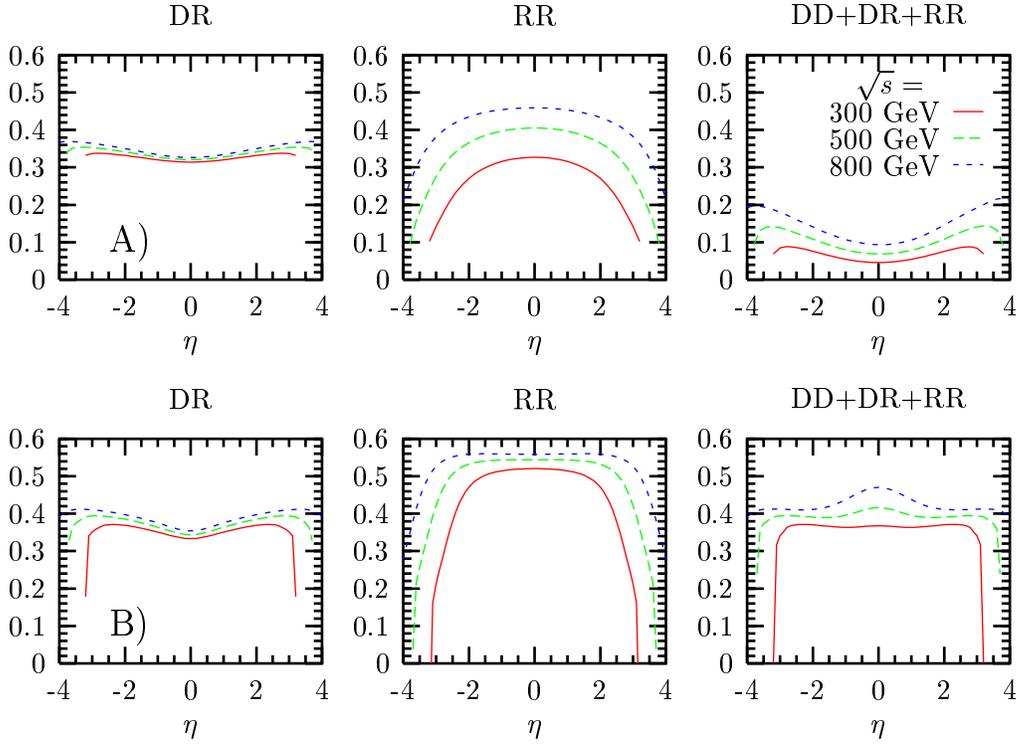}
\vskip -17cm
\caption{As in Fig. 6,  for the FFNS scheme.}
\end{figure}

\clearpage

\begin{figure}
\vskip -5cm
\hskip -4cm
\psfig{file=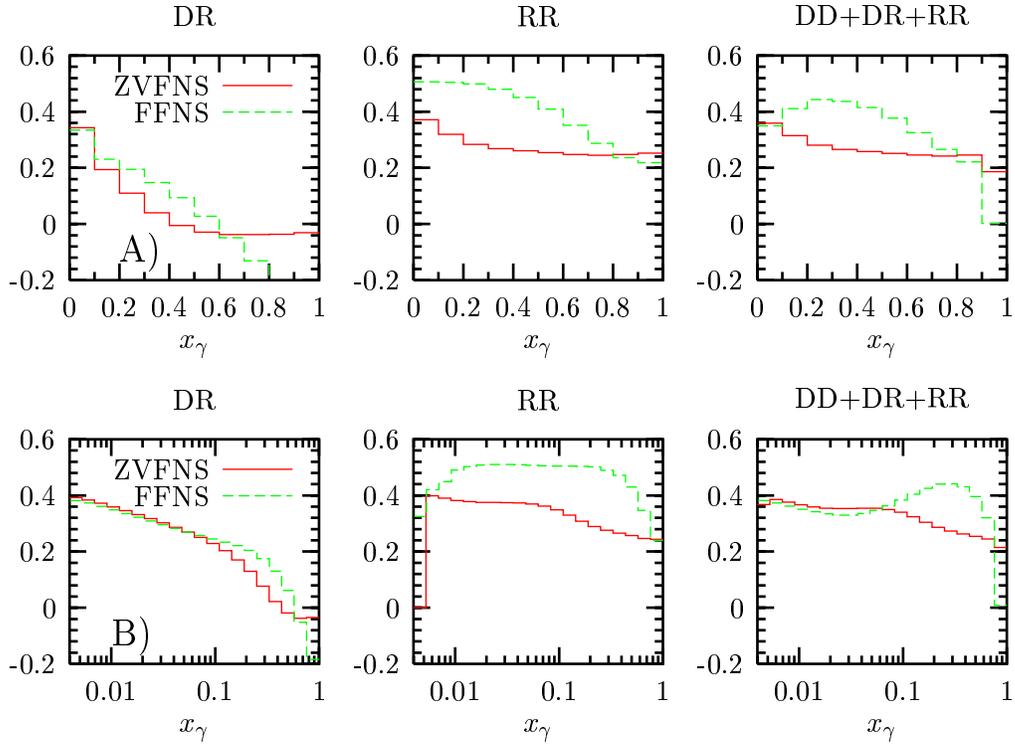}
\vskip -17.5cm
\caption{
The ratio {\large $\frac{GRV-SAS1d}{GRV}$}
of the cross-section {\Large $\frac{d\sigma}{dx_{\gamma}}$}
$(e^+e^- \rightarrow e^+e^- c/\bar{c} X)$ as a function of $x_{\gamma}$
at $\sqrt{s}=500$ GeV for $p_T>10$ GeV and $|\eta|<2$. A comparison is shown 
of the  two schemes  for a LASER spectrum on A) linear and  B) logarithmic 
$x_{\gamma}$ scale.}
\end{figure}

\begin{figure}
\vskip -3.5cm
\hskip -4cm
\psfig{file=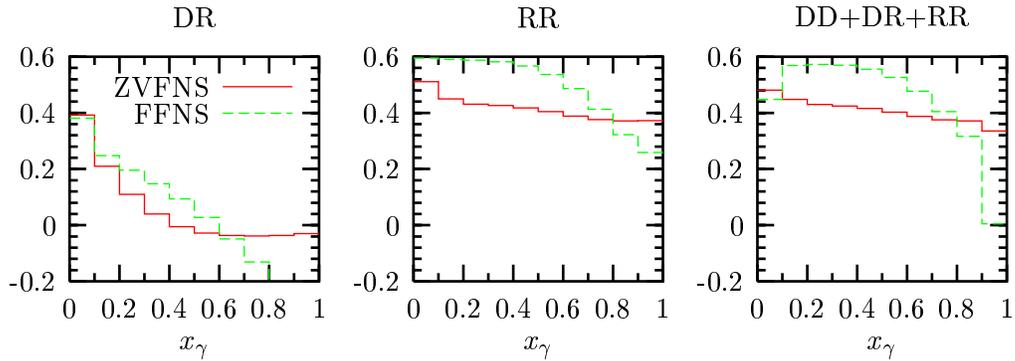}
\vskip -22.5cm
\caption{As in Fig. 8, for $p_T>5$ GeV and linear scale only.}
\end{figure}

\clearpage

\begin{table}[h]
\begin{center}
\begin{tabular}{|c|c|c|c|c|c|c|c|c|}
\hline
    $p_T>10$ GeV            & DD & \multicolumn{2}{c|}{DR} & \multicolumn{5}{c|}{RR} \\
\hline
        & $\gamma + \gamma$ & $g+\gamma$ & $Q+\gamma$ & $g+g$ & 
$q+\bar q$ & $Q+\bar Q$, $Q+Q$& $Q+q$ & 
$Q(\bar Q)+g$ \\
\hline
 $\sqrt{s}=300\: GeV$ & 3.1 & 40 & 5.6 & 0.34 & 5.5 & 1.3 & 17 & 33 \\
\hline
 $\sqrt{s}=500\: GeV$ & 1.1 & 43 & 4.6 & 0.37 & 6.3 & 3.5 & 20 & 46 \\
\hline
 $\sqrt{s}=800\: GeV$ & 0.43 & 42 & 3.8 & 0.41 & 7.5 & 7.9 & 24 & 62 \\
\hline
\end{tabular}
\end{center}
\caption{Cross-sections (in $pb$) for charm  quark production 
with $p_T>10$ GeV and $|\eta|<2$ ($0.6<y<0.83$), calculated in the 
massless ZVFNS scheme with the GRV parton parametrization.}
\end{table}

\begin{table}[h]
\begin{center}
\begin{tabular}{|c|c|c|c|c|}
\hline
 $p_T>10$ GeV & \multicolumn{3}{c|}{ZVFNS} & FFNS \\
\hline
                      & $\sigma(g)/\sigma$ & $\sigma(c)/\sigma$ & 
 $ \sigma(g+c)/\sigma$ & $ \sigma(g)/\sigma$ \\
\hline
 $\sqrt{s}=300\: GeV$ & 0.70 & 0.58 & 0.97 & 0.92 \\
\hline
 $\sqrt{s}=500\: GeV$ & 0.74 & 0.61 & 0.99 & 0.97 \\
\hline
 $\sqrt{s}=800\: GeV$ & 0.76 & 0.66 & 0.99 & 0.98 \\
\hline
\end{tabular}
\end{center}
\caption{An estimate of the partonic  contributions to the charm quark 
production  with $p_T>10$ GeV and $|\eta|<2$ ($0.6<y<0.83$), calculated in 
both the massless ZVFNS and massive FFNS schemes with the GRV parton 
parametrization. The sum of the contributions which involve 
gluons ($\sigma(g)$), charm quarks ($\sigma(c)$) or both ($\sigma(g+c)$)
in the initial state are compared to the total  cross-section $\sigma$.}
\end{table}

\begin{table}[hb]
\begin{center}
\begin{tabular}{|c|c|c|c|c|c|c|}
\hline
                      & \multicolumn{4}{c|}{$p_T>10$ GeV} & \multicolumn{2}{c|}{$p_T>5$ GeV} \\
\hline
                      & \multicolumn{2}{c|}{$c/\bar{c}$} & \multicolumn{2}{c|}{$b/\bar{b}$} & \multicolumn{2}{c|}{$c/\bar{c}$} \\
\hline
                      &  LC  &  PC   &  LC   &  PC    & LC    & PC \\
\hline
 $\sqrt{s}=300\: GeV$ & 4.0 (5.4) & 42 (110)  & 0.27 (0.22)  & 8.4 (23)  & 25 (39)  & 170 (490)  \\
\hline
 $\sqrt{s}=500\: GeV$ & 6.0 (8.6) & 45 (120)  & 0.46 (0.94) & 10 (32)  & 35 (58)  & 180 (600)  \\
\hline
 $\sqrt{s}=800\: GeV$ & 8.2 (13) & 47 (150)  & 0.73 (1.6)  & 13 (47)  & 47 (83)  & 200 (770)  \\
\hline
\end{tabular}
\end{center}
\caption{Cross-sections (in $pb$) for heavy quark production with $|\eta|<2$,
at a LC and a PC ($0.6<y<0.83$), calculated in the massive FFNS (ZVFNS) scheme
with the GRV parton parametrization.}
\end{table}

\begin{table}[h]
\begin{center}
\begin{tabular}{|c|c|c|c|c|c|c|}
\hline
                      & \multicolumn{4}{c|}{$p_T>10$ GeV} & \multicolumn{2}{c|}{$p_T>5$ GeV} \\
\hline
                      & \multicolumn{2}{c|}{$c/\bar{c}$} & \multicolumn{2}{c|}{$b/\bar{b}$} & \multicolumn{2}{c|}{$c/\bar{c}$} \\
\hline
                      &  LC   &  PC    &  LC   &  PC    &  LC   & PC \\
\hline
 $\sqrt{s}=300\: GeV$ & 1.2  & 3.9 & 0.08 & 0.80 & 7.5 & 16 \\
\hline
 $\sqrt{s}=500\: GeV$ & 1.8  & 4.1 & 0.14 & 0.97 & 11 & 17 \\
\hline
 $\sqrt{s}=800\: GeV$ & 2.5 & 4.4 & 0.22 & 1.2 & 14 & 19 \\
\hline
\end{tabular}
\end{center}
\caption{Event rates (in units of $10^6$) for heavy quark production with 
$|\eta|<2$ at a LC and a PC ($0.6<y<0.83$), calculated in the massive FFNS 
scheme with the GRV parton parametrization. The $e^+e^-$ LC luminosity was 
assumed to be 300 fb$^{-1}$.}
\end{table}

\end{document}